\documentstyle[seceq,amssymb,epsfig]{ptptex}

\notypesetlogo

\def\kreis{\raise0.85pt\hbox{$\scriptstyle\bigcirc$}}
\def\vollk{\lower0.85pt\hbox{\Large $\bullet$}}

\def\half{\raise0.85pt\hbox{$\scriptstyle\frac{1}{2}$}}

\title{
\vspace{-1.9cm}
\begin{flushleft}
{\normalsize DESY 98-065} \hfill\\
{\normalsize May 1998} 
\end{flushleft}
\vspace{0.4cm}
Chiral Fermions on the Lattice\,\thanks{Talk given by G. Schierholz
at {\it Yukawa International Seminar on Non-Perturbative QCD: Structure of
the QCD Vacuum (YKIS97)}, Kyoto, December 1997}
\\[0.2em]} 

\author{V. {\sc Bornyakov}$^1$, G. {\sc Schierholz}$^{2,3}$
and A. {\sc Thimm}$^{2,4}$\\[0.4em]}

\inst{$^1$Institute for High Energy Physics IHEP, RU-142284 Protvino, 
      Russia \\[0.4em]
      $^2$Deutsches Elektronen-Synchrotron DESY, D-15735 Zeuthen, 
      Germany \\[0.4em]
      $^3$Deutsches Elektronen Synchrotron DESY, D-22603 Hamburg,
      Germany \\[0.4em]
      $^4$ Institut f\"ur Theoretische Physik, Freie Universit\"at
      Berlin, D-14195 Berlin, Germany}

\abst{A recently proposed method for regularizing chiral gauge 
theories non-perturbatively is discussed in detail. The result is an 
effective action which can be computed from the lattice gauge field, 
and which is suited for numerical simulations.}

\begin{document}

\maketitle

\section{Introduction}

Formulating chiral gauge theories on the lattice is one of the
great challenges in particle physics. The solution of this problem would
lead to a new, qualitative and quantitative understanding of these
theories. 

To pay tribute to the topological nature of the problem, it has been
suggested~\cite{GS} to discretize the gauge fields only and 
consider the fermions in the continuum. By this we mean the following. 
One starts from a lattice with lattice spacing $a$. We call this lattice the
original lattice. This is also the lattice on which the simulations of the 
gauge fields will be done. Then one constructs a finer
lattice with lattice spacing $a_f$. On this lattice one places the
fermions. We shall assume Wilson fermions to remove the doublers. Before one 
can state the action on the fine lattice, one has to 
extrapolate the 
gauge fields to the interior of the original lattice. A suitable extrapolation
was given in ref.~2. The effective fermionic action is then derived in
a two-step procedure. In the first step one computes the lattice effective
action in the limit $a_f \rightarrow 0$, while keeping $a$ fixed. This action
is generally not invariant under chiral gauge transformations.
Chiral gauge invariance can, however, be restored by adding a few local 
bosonic counterterms to the action. The second step then is to determine the 
counterterms. This can be done perturbatively. 
The resulting action is a non-local function of the gauge fields on the
original lattice. One might think that the action 
will be too complicated to be of any use. But we shall 
see that this is not the case. On the contrary: a simple effective action 
emerges which lends itself to numerical simulations. 
For similar ideas see ref.~3.

In this talk we shall test our idea in the chiral Schwinger model. The 
Schwinger model is particularly well suited for this task because a lot is
known analytically. The talk is organized as follows. 
In sec.~2 we
present some selected analytic results. In sec.~3 we derive
the effective fermionic action including the counterterms. In sec.~4 we
then show that the resulting action is gauge invariant, and we compare 
numerical and analytic results. Finally, in sec.~5 we conlude.

\section{Analytic Results}

We shall mainly be concerned with the fermionic action. In the vector 
Schwinger model it reads
\begin{equation}
S  =  \sum_{\alpha} \int \mbox{d}^2x \, \bar{\psi}_\alpha(x) 
\not{\!\!D}(e_\alpha A) \psi_\alpha(x),
\end{equation}
where the sum is over fermion flavors, $e_\alpha$ is the charge of the 
fermion in units of $e$, and
\begin{equation}
D_\mu(A) = \partial_\mu + \mbox{i}A_\mu.
\end{equation}
Note
that $e_\alpha$ is dimensionless, but $e$ and $A_\mu$ are dimensionful.
The model has a topological charge
\begin{equation}
Q = \frac{1}{4\pi} \int \mbox{d}^2x \, \epsilon_{\mu\nu} F_{\mu\nu}(x) 
\in {\Bbb Z}.
\end{equation}
The Schwinger model has been solved analytically in 
${\Bbb R}^2\,$~\cite{Schwinger}, 
on the sphere ${\Bbb S}^2\,$~\cite{Jaye}, as well as on the torus 
${\Bbb T}^2\,$~\cite{Joos,Wipf,Azakov}.

In the chiral Schwinger model the fermionic action reads
\begin{equation}
S =  \sum_{\alpha} \int \mbox{d}^2x \, 
\bar{\psi}_\alpha(x) 
\not{\!\!D}^{\epsilon_\alpha}(e_\alpha A) \psi_\alpha(x),
\label{caction}
\end{equation}
where $\epsilon_\alpha = \pm 1$ is the chirality, and 
\begin{equation}
D^{\epsilon_\alpha}_\mu(A) = \partial_\mu 
+ \mbox{i} A_\mu P_{\epsilon_\alpha},
\end{equation}
with 
\begin{equation}
P_{\epsilon_\alpha} = \frac{1}{2}(1 + \epsilon_\alpha \gamma_5), 
\gamma_5 = \mbox{i}\gamma_1\gamma_2 = \left( \begin{array}{rr} -1 & 0\\
0 & 1 \end{array} \right). 
\end{equation}
In eq.~(\ref{caction}) the sum is over flavors and chiralities.
The action (\ref{caction}) is invariant under chiral gauge transformations
$g^\alpha = \exp(\mbox{i}e_\alpha h P_{\epsilon_\alpha})$:
\begin{eqnarray}
 \psi_\alpha &\rightarrow&  \psi^g_\alpha = g^\alpha \psi_\alpha, \nonumber \\
\bar{\psi}_\alpha &\rightarrow& \bar{\psi}^g_\alpha = 
\bar{\psi}_\alpha g^\alpha, \label{CGT} \\
A_\mu &\rightarrow& A^g_\mu = A_\mu - \partial_\mu h. \nonumber
\end{eqnarray}
The anomaly cancelling condition is
\begin{equation}
\sum_{\alpha} \epsilon_\alpha e^2_\alpha = 0.
\end{equation}

The effective fermionic action is defined by
\begin{equation}
\exp(-W) = \int {\cal D}\bar{\psi} {\cal D}\psi 
\exp(-S).
\end{equation}
In ${\Bbb R}^2$, in a background of trivial topology $Q = 0$, the effective 
action is known analytically both for the vector and the chiral 
model~\cite{Schwinger,JR,DL}. 
After subtracting the divergent contribution arising from the free 
fermions, we obtain
\begin{equation}
W(A) - W_0 = \sum_{\alpha} \frac{e^2_\alpha}{8 \pi} \int \mbox{d}^2x A_\mu(x) 
\left[a \delta_{\mu\nu} - (\partial_\mu + 
\mbox{i}\epsilon_\alpha\widetilde{\partial}_\mu)
\frac{1}{\square} (\partial_\nu + 
\mbox{i}\epsilon_\alpha\widetilde{\partial}_\nu)\right] 
A_\nu(x),
\label{W}
\end{equation}
where $W_0 = W(0)$, 
\begin{equation}
\widetilde{\partial}_\mu = \epsilon_{\mu\nu} \partial_\nu
\end{equation}
and $a$ (not to be confused with the lattice spacing $a$) reflects a 
regularization ambiguity~\cite{JR,Jackiw}. The 
one-loop perturbative result corresponds to (\ref{W}) with $a = 1$. 
For $a = 1$ the effective action is 
invariant under chiral gauge transformations (\ref{CGT}). It follows that
\begin{equation}
\mbox{Re} W(A) = \frac{1}{2} \left(W_V(A) + W_0\right),
\end{equation}
where $W_V$ is the effective action of the corresponding vector model.
Moreover, we find
\begin{equation}
\mbox{Im} W(A) = 0
\label{imagr}
\end{equation}
in the anomaly-free model. For an extension of the results to non-trivial
topology see ref.~12.

On the $T_1 \times T_2$ torus the gauge field can be decomposed 
according to
\begin{equation}
A_\mu(x) = \frac{2 \pi}{T_\mu} t_\mu + \partial_\mu h(x) 
+ \widetilde{\partial}_\mu f(x) + C^Q_\mu(x),
\end{equation}
where $t_\mu$ is the zero-momentum component of the gauge field, called
toron, $\partial_\mu h$ represents the pure gauge degrees of freedom, and 
$\widetilde{\partial}_\mu f$ and $C^Q_\mu$ are the proper dynamical 
components of zero and
non-zero topological charge, respectively. Note, however, that this decomposition
is not unique with respect to large gauge transformations~\cite{Azakov}. 
%The vector model is invariant under the gauge transformation
%\begin{equation}
%g^\alpha = \exp\left(\mbox{i}\frac{2 \pi}{T_\mu} n_\mu x_\mu\right), \; 
%n_\mu \in {\Bbb Z}
%\label{SH}
%\end{equation}
%which corresponds to the shift $t_\mu \rightarrow t_\mu + n_\mu$. The same
%holds for the anomaly-free chiral model if we use anti-periodic fermionic
%boundary conditions
%\begin{equation}
%\psi(x + m T_1 + n T_2) = \exp(\mbox{i} \pi (m + n)) \psi(x).
%\end{equation}
%This will become clear later on.
%So in these two cases we can restrict $t_\mu$ to $-1 < t_\mu \leq 1$.
Let us now consider the sector of zero topological charge $Q = 0$. For the 
gauge field we then may write
\begin{equation}
A_\mu(x) = a_\mu + b_\mu(x),
\end{equation}
where $a_\mu = (2 \pi/T_\mu)\: t_\mu$ and 
$b_\mu(x) =\widetilde{\partial}_\mu f(x) + \partial_\mu h(x)$. We will show 
now that the effective action factorizes:
\begin{equation}
W(A) = W(a) + W(b).
\label{factor}
\end{equation}
We define
\begin{equation}
A^\tau_\mu(x) = \tau A_\mu(x),
\end{equation}
with $0 \leq \tau \leq 1$. The effective action is then given by~\cite{AG}
\begin{eqnarray}
W(A) - W_0 &=& -\frac{1}{2} \sum_{\alpha} \int^1_0 \mbox{d}\tau 
\mbox{Tr} (\frac{\mbox{d}}{\mbox{d}\tau} 
\not{\!\!D}^{\epsilon_\alpha}(e_\alpha A^\tau))
\not{\!\!D}(e_\alpha A^\tau)^{-1} \\
           &=& -\frac{\mbox{i}}{2} \sum_{\alpha} e_\alpha \int^1_0 
           \mbox{d}\tau \mbox{Tr} \not{\!\!A} \: P_{\epsilon_\alpha} 
           \not{\!\!D}(e_\alpha A^\tau)^{-1}.
\end{eqnarray}
The particular feature of this expression is that it
involves the covariant Dirac operator of the vector model
only. To make the expression well defined, we use point splitting
regularization:
\begin{eqnarray}
W(A) - W_0 &=& - \lim_{\epsilon \rightarrow 0} 
\frac{\mbox{i}}{4} \sum_{\alpha} \left\{ e_\alpha 
\int^1_0 \mbox{d}\tau \int\mbox{d}^2x \not{\!\!A}(x) P_{\epsilon_\alpha} 
G(e_\alpha A^\tau \mid x,x+\epsilon)\right. \nonumber \\
           & & \left. \times \exp(\mbox{i}\int^x_{x+\epsilon} 
\mbox{d}z_\mu A^\tau_\mu(z)) + (\epsilon \leftrightarrow - \epsilon)\right\},
\label{alvarez}
\end{eqnarray}
where the fermion propagator $G(A \mid x,y)$ is a solution of the equation
\begin{equation}
\not{\!\!D}(A) G(A \mid x,y) = \delta(x - y)
\label{prop}
\end{equation}
with $\delta(x-y)$ being the torus $\delta$ function.
If $G(A \mid x,0)$ is a solution of (\ref{prop}) with periodic boundary 
conditions,
\begin{equation}
G(A \mid x + m T_1 + n T_2,0) = G(A \mid x,0),
\end{equation}
then
\begin{equation}
G^c(A \mid x,0) = \exp(\mbox{i} cx) G(A + c \mid x,0) 
\end{equation}
is a solution with boundary conditions
\begin{equation}
G^c(A \mid x + m T_1 + n T_2,0) = \exp(\mbox{i}mc_1T_1 + \mbox{i}nc_2T_2) 
G^c(A \mid x,0).
\end{equation}
For $c_\mu = \pi/T_\mu$ this corresponds to anti-periodic boundary conditions.
The propagator can be written 
\begin{equation}
G(A \mid x,y) = \exp(\mbox{i} h(x) + \gamma_5 f(x))\: G(a \mid x,y)\:
                \exp(-\mbox{i} h(y) + \gamma_5 f(y)). 
\label{fac}
\end{equation}
At short distances $|x-y|$ we have~\cite{Azakov}
\begin{equation}
G(a \mid x,y) = \exp(-\mbox{i} a (x-y))
\left(G_0(x,y) - \frac{1}{2\pi} K(a) \right) 
+ O(|x-y|),
\label{short}
\end{equation}
with $G_0(x,y) = G(0 \mid x,y)$ and
\begin{equation}
K(a) = \left( \begin{array}{cc} 0 & {\displaystyle -\mbox{i}a_1 + 
\frac{1}{T_1} \frac{\Theta^\prime_1(t_-)}{\Theta_1(t_-)}} \\
{\displaystyle -\mbox{i}a_1 + 
\frac{1}{T_1} \frac{\Theta^\prime_1(t_+)}{\Theta_1(t_+)}}  & 0 \end{array} 
\right), 
\end{equation}
where $\Theta_1(x) \equiv \Theta_1(x \mid \mbox{i} T_2/T_1)$ is the Jacobi
function and $t_\pm = (T_2/2\pi)(a_2 \pm \mbox{i} a_1)$.
From eqs.~(\ref{fac}) and (\ref{short}) we see already that the effective
action factorizes. It is straightforward now to evaluate 
eq.~(\ref{alvarez}). Using standard techniques we obtain for general 
boundary conditions
\begin{eqnarray}
W(A) - W_0 &=& \sum_{\alpha} \left\{\frac{T_1T_2}{4\pi} e_\alpha a_1 
               (e_\alpha a_1 + 2c_1) - \ln |\Theta_1(e_\alpha t_+ + c_+)|
               + \ln |\Theta_1(c_+)|\right. \nonumber \\
           &-& \left. \mbox{i} \epsilon_\alpha \left[\frac{T_1T_2}{4\pi} 
               e_\alpha a_2 (e_\alpha a_1 + 2c_1) + 
               \arg \Theta_1(e_\alpha t_+ + c_+) -
               \arg \Theta_1(c_+)\right]\right\} \nonumber \\
           &+& \sum_{\alpha} \frac{e^2_\alpha}{8 \pi} \int 
               \mbox{d}^2x b_\mu 
               \left[\delta_{\mu\nu} - (\partial_\mu + \mbox{i}\epsilon_\alpha
               \widetilde{\partial}_\mu)\frac{1}{\square} 
               (\partial_\nu + \mbox{i}\epsilon_\alpha\widetilde{\partial}_\nu)
               \right]b_\nu,
\label{analy}
\end{eqnarray}
where $c_+ = (T_2/2\pi)(c_2+\mbox{i}c_1)$. This proves eq.~(\ref{factor}).
It follows that
\begin{equation}
\mbox{Re} W(A) = \frac{1}{2} \left(W_V(A) + W_0\right),
\end{equation}
as in ${\Bbb R}^2$. In contrast to the previous result (\ref{imagr}), 
the imaginary part of the effective action is no longer zero in
the anomaly-free model. The reason for that is the toron 
field contribution~\cite{note}:
\begin{equation}
\mbox{Im} W(A) = - \sum_{\alpha} \epsilon_\alpha 
\left[\frac{T_1T_2}{4\pi} 
               e_\alpha a_2 (e_\alpha a_1 + 2c_1) + 
               \arg \Theta_1(e_\alpha t_+ + c_+) -
               \arg \Theta_1(c_+)\right].
\label{imag}
\end{equation}

A particular choice of $C^Q_\mu(x)$ is~\cite{Joos,Azakov}
\begin{equation}
C^Q_\mu(x) = - \frac{\pi Q}{T_1 T_2} \epsilon_{\mu\nu} x_\nu.
\label{charge}
\end{equation} 
%which gives rise to a constant field strength tensor $F_{\mu\nu} = 
%(e\pi Q/T_1T_2) \epsilon_{\mu\nu}$, and which fulfills the Landau gauge
%condition $\partial_\mu C^Q_\mu = 0$. The gauge field (\ref{charge}) is not
%periodic but requires the following periodicity conditions to be
%fulfilled:
%\begin{eqnarray}
%\psi(x + T_\nu \hat{\nu}) &=& \Lambda_\nu(x) \psi(x), \nonumber \\
%\bar{\psi}(x + T_\nu \hat{\nu}) &=& \bar{\psi(x)}\Lambda^{-1}_\nu(x), \\
%A_\mu(x + T_\nu \hat{\nu}) &=& A_\mu(x) - \mbox{i} \Lambda^{-1}_\nu(x) 
%\partial_\mu \Lambda_\nu(x),
%\end{eqnarray}
%where
%\begin{equation}
%\Lambda_\mu(x) = \exp\left(\mbox{i} \pi Q \epsilon_{\mu\nu} 
%\frac{x_\nu}{T_\nu}\right)
%\end{equation}
%which satisfies the cocycle condition.
The Dirac operator
\begin{equation}
\not{\!\!D}^{Q} = \not{\!\partial} + \mbox{i} e_\alpha \not{\!\!C}^{Q}
\end{equation}
has $|Q|$ zero mode solutions:
\begin{equation}
\not{\!\!D}^{Q} \chi^{Q}_l = 0, \; l=1, \cdots , |Q|.
\end{equation}
%The eigenfunctions are of the form
%\begin{equation}
%\chi^{Q}_l = \left\{ \begin{array}{ll}
%      \left(\begin{array}{c} \chi_1 \\ 0 \end{array}\right) & Q > 0, \\
%      & \\
%      \left(\begin{array}{c} 0 \\ \chi_2 \end{array}\right) & Q < 0. 
%                       \end{array} \right.
%\end{equation}
%They are known analytically~\cite{Azakov}. 
Each eigenfunction has a definite chirality:
\begin{equation}
\chi^{Q\dagger}_l \gamma_5 \chi^{Q}_l = \left\{ \begin{array}{ll}
      + 1 & Q > 0, \\
      - 1 & Q < 0. 
\end{array} \right.
\end{equation}
The number of zero modes $n_+$ ($n_-$) with positive (negative) 
chirality is then given by
\begin{eqnarray}
n_+ &=& Q \, \theta(Q), \nonumber \\
n_- &=& |Q| \, \theta(-Q),
\label{zero} 
\end{eqnarray}
which is the index theorem (in two dimensions).
Accordingly, the chiral Dirac operator $\not{\!\!\!D}^{Q\,\epsilon_\alpha}$ 
has $|Q|$ zero modes of chirality $\epsilon_\alpha$ if and only if
$\epsilon_\alpha = \mbox{sign}(Q)$. The eigenfunctions are the same as in the
vector case.
Later on we will use the expression
\begin{equation}
C^Q_\mu(x) = 2\pi Q \epsilon_{\mu\nu} \partial_\nu G(x), 
\; x \in \dot{\Bbb T}^2,
\label{gaugedef}
\end{equation}
where $\dot{\Bbb T}^2$ is the torus with the point $x = 0$ removed, 
and $G(x)$ is the inverse Laplacian on the torus satisfying the equation
\begin{equation}
- \square G(x) = \delta(x) - \frac{1}{T_1T_2}.
\end{equation}
In analytic form~\cite{Azakov}
\begin{equation}
G(x) = \frac{1}{2} \frac{x^2_2}{T_1T_2} - 
\frac{1}{2\pi} \mbox{Re} \ln 
\left(\frac{\Theta_1(z)}{\eta(\mbox{i}T_2/T_1)}\right),
\end{equation}
where $z=(x_1 + \mbox{i}x_2)/T_1$ with $\Theta^\prime_1(0) = 
2\pi\eta(\mbox{i}T_2/T_1)^3$.
The gauge field (\ref{gaugedef}) is periodic, and it is related to the former 
expression (\ref{charge}) by the gauge transformation
\begin{equation}
h(x) = Q\, \mbox{Re}\: \mbox{i} \ln 
\left(\frac{\Theta_1(z)}{\eta(\mbox{i}T_2/T_1)}\right)
+\frac{\pi Q}{T_1T_2} x_1x_2.
\end{equation}

\section{Effective Fermionic Action}

%Before one can do this, one has to extrapolate the gauge fields to the
%interior of the original lattice. That means one has to derive continuum
%gauge fields from lattice gauge fields. The continuum gauge fields must
%fulfill the following constraints: (i) The parallel transporters derived
%from the continuum gauge field agree with those of the original lattice
%gauge field. (ii) A lattice gauge transformation results in a gauge
%transformation of the continuum gauge field, i.e. the extrapolation is
%gauge covariant. (iii) The continuum gauge field is a connection in the
%fiber bundle~\cite{?} which has been constructed from the lattice gauge
%field. This guarantees that one obtains the same value for the
%topological charge, independent of whether one computes it from the
%transition functions or by integrating the Chern-Pontryagin density.
%Such an extrapolation was given in~\cite{GSW}.

We shall first compute the lattice effective fermionic 
action and then determine the counterterms. We start from a 
$L_1 \times L_2$ lattice with lattice spacing $a$.
From this lattice we construct
a fine $L^f_1 \times L^f_2$ lattice with lattice spacing $a_f$. In
practice $a_f = a/N, N \equiv L^f_\mu/L_\mu \in {\Bbb N}$.
Let $n_\mu$, $1 \leq n_\mu \leq L^f_\mu$, denote 
the points on the fine lattice. The action for one species of fermion
with charge $e_\alpha$ and chirality $\epsilon_\alpha$ is given by
\begin{eqnarray}
S_{\epsilon_\alpha} &=& \frac{1}{2a_f} \sum_{n,\mu} \left\{\bar{\psi}(n) 
  \gamma_\mu [(P_{-\epsilon_\alpha} + P_{\epsilon_\alpha}
  (U^f_\mu(n))^{e_\alpha} 
  \psi(n+\hat{\mu})\right. \nonumber \\
 &-& \left.(P_{-\epsilon_\alpha} + P_{\epsilon_\alpha}
  (U^{f}_\mu(n-\hat{\mu}))^{e_\alpha \dagger} 
  \psi(n-\hat{\mu})]\right\} \label{LA} \\
 &+& S_{W \epsilon_\alpha}(U^f), \nonumber
\end{eqnarray}
where $\hat{\mu}$ is a unit vector in $\mu$-direction on 
the fine lattice,
and $S_{W \epsilon_\alpha}(U^f)$ is the Wilson term. 
On the fine lattice the link variables are
\begin{equation}
U^f_\mu(n) \equiv \exp(\mbox{i}\theta^f_\mu(n)) = 
\exp\left(\mbox{i} a_f \int^{n+\hat{\mu}}_n \mbox{d}z_\mu A_\mu(z)\right),
\label{link}
\end{equation}
where $A_\mu$ is the continuum gauge field obtained by extrapolation.

\begin{figure}[t]
\vspace{-4.0cm}
\begin{centering}
\epsfig{figure=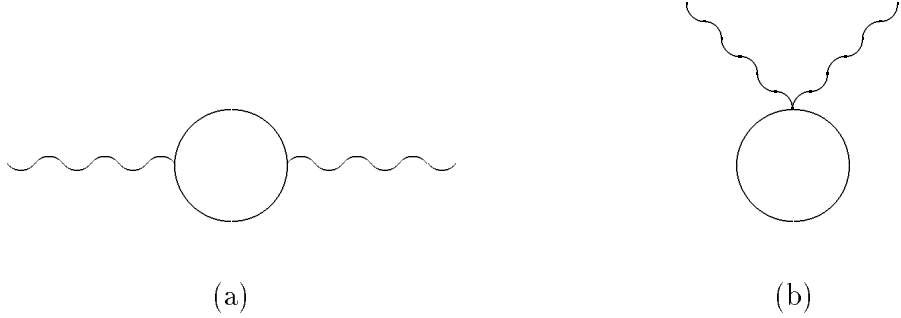,height=15.0cm,width=15.0cm}
\vspace{-4.5cm}
\caption{Fermion loop diagrams contributing to the effective action.}
\end{centering}
%\vspace{-2.0cm} 
\end{figure}

The lattice effective action is given by
\begin{equation}
\exp(-W_{\epsilon_\alpha}) = \int {\cal D}\bar{\psi}{\cal D}\psi
\exp(-S_{\epsilon_\alpha}).
\end{equation}
Due to the presence of the Wilson term, this action is not 
invariant under chiral gauge transformations.
Classically the Wilson term vanishes to order $a_f$. However, in fermion 
loops as shown in fig.~1, which receive contributions from loop momenta of 
the order of the cut-off $\approx \pi/a_f$, the Wilson term
will in general give a finite contribution to the effective action. 
Because the effective action refers to classical background gauge fields, 
we need not consider diagrams with internal gauge boson lines. 
In the limit $a_f \rightarrow 0$ the non-gauge invariant contribution of
the loop integrals 
is expected to contract to a multi-gauge boson amplitude, so that chiral
gauge invariance may be restored by adding a few local bosonic counterterms
to the action. The advantage of this method, as opposed to 
the Rome approach~\cite{Rome}, is that the counterterms can be computed 
perturbatively, and no (non-perturbative) fine-tuning of the coefficients is 
required. We write
\begin{equation}
W^\Sigma_{\epsilon_\alpha} = W_{\epsilon_\alpha} + C,
\label{WC}
\end{equation}
where $C$ denotes the counterterm. The task is then to determine $C$,
so that 
\begin{equation}
\widehat{W}_{\epsilon_\alpha} = \lim_{a_f \rightarrow 0} 
W^\Sigma_{\epsilon_\alpha}
\label{effective}
\end{equation}
is invariant under chiral gauge transformations in the anomaly-free model.

For the Wilson term one has several choices. We shall take 
\begin{eqnarray}
S_{W \epsilon_\alpha}(U^f) &=& - \frac{r}{2} \sum_{n,\mu} \bar{\psi}(n)
D^{\epsilon_\alpha +}_\mu D^{\epsilon_\alpha -}_\mu \psi(n)
\nonumber \\
&=& \frac{r}{2 a_f}\sum_{n,\mu} \bar{\psi}(n)\{2\psi(n) 
- [P_{-\epsilon_\alpha} + 
P_{\epsilon_\alpha}(U^f_\mu(n))^{e_\alpha}]\psi(n+\hat{\mu}) \nonumber \\
& & - [P_{-\epsilon_\alpha} + 
P_{\epsilon_\alpha}(U^f_\mu(n-\hat{\mu}))^{e_\alpha \dagger}]
\psi(n-\hat{\mu})\}, \label{PG} 
\end{eqnarray}
where the superscript $+$ ($-$) means forward (backward) derivative.
%The Wilson term has the property
%\begin{equation}
%S_{W +}(U^f) + S_{W -}(U^f) = S_W(U^f) + S_W(0),
%\end{equation}
%where $S_W$ is the gauged Wilson term from the vector model. This leads to
%\begin{equation}
%S_+ + S_- = S_V + S_0,
%\end{equation}
%where $S_V$ ($S_0$) is the action of the vector (free) theory.
%Any other choice of Wilson term does not have this property. 
This Wilson term has two important properties which, e.g., the ungauged and 
the gauged Wilson terms of the vector model do not have.  

In order that the lattice theory is in the same universality class as the 
classically defined theory, it must obey the index theorem. 
The lattice fermionic action can be written in the matrix form
\begin{equation}
\bar{\psi}\left(\not{\!\!D}^{\epsilon_\alpha} - \frac{r}{2} 
M^{\epsilon_\alpha} \right)\psi.
\end{equation}
The Dirac operator $\not{\!\!D}^{\epsilon_\alpha}$ is a finite
matrix now. Thus $\not{\!\!D}^{\epsilon_\alpha}$ and
$\not{\!\!D}^{\epsilon_\alpha \dagger}$ have the same number of zero modes.
It is furthermore easy to see that
\begin{equation}
\not{\!\!D}^{\epsilon_\alpha \dagger} = \not{\!\!D}^{-\epsilon_\alpha}.
\end{equation}
This tells us that the chiral lattice Dirac operator has the 
same number of right-handed and left-handed zero modes, thus violating the 
index theorem (\ref{zero}). The situation is different if we include the
Wilson term. A necessary requirement for the index theorem to be valid on
the lattice is
\begin{equation}
M^{\epsilon_\alpha \dagger} \neq M^{-\epsilon_\alpha}.
\label{M}
\end{equation}
The Wilson term (\ref{PG}) fulfills this requirement, unlike the ungauged 
and the
gauged Wilson term of the vector model. In ref.~17 we will show that
the Wilson term (\ref{PG}) in fact fulfills the index theorem.

The Wilson term couples right- and left-handed fermions. 
In order that the theory is chiral, the ungauged fermion must 
decouple in the limit $a_f \rightarrow 0$. This is the case
if the action is invariant
under the transformations~\cite{GP}
\begin{eqnarray}
\psi_{-\epsilon_\alpha} \equiv P_{-\epsilon_\alpha}\psi 
& \rightarrow & \psi_{-\epsilon_\alpha} + \eta, \label{S1} \\
\bar{\psi}_{-\epsilon_\alpha} \equiv \; \bar{\psi} P_{\epsilon_\alpha} \: 
& \rightarrow & \bar{\psi}_{-\epsilon_\alpha} + \bar{\eta}, \label{S2}
\end{eqnarray}
where $\eta$, $\bar{\eta}$ are constant Grassmann variables. We denote the 
fermion propagator corresponding to the action (\ref{LA}) in momentum
space by $G_{\epsilon_\alpha}(p)$. If the action is invariant under the 
transformation (\ref{S1}), we find
\begin{eqnarray}
G^{-1}_{\epsilon_\alpha}(p) &=& \left(\frac{\mbox{i}}{a_f} \gamma_\mu 
f^{\epsilon_\alpha}_\mu(p) + m^{\epsilon_\alpha}(p)\right) 
P_{\epsilon_\alpha} \nonumber \\
& & + 
\left(\frac{\mbox{i}}{a_f} \gamma_\mu \sin(a_fp_\mu) 
+ \frac{r}{a_f}\sum_\mu[1-\cos(a_fp_\mu)]\right)P_{-\epsilon_\alpha}, 
\label{shift1}
\end{eqnarray}
and if the action is invariant under the transformation (\ref{S2}), 
we find 
\begin{eqnarray}
G^{-1}_{\epsilon_\alpha}(p) &=& \left(\frac{\mbox{i}}{a_f} \gamma_\mu 
f^{\epsilon_\alpha}_\mu(p) + \frac{r}{a_f}\sum_\mu[1-\cos(a_fp_\mu)]\right) 
P_{\epsilon_\alpha} \nonumber \\
& & + 
\left(\frac{\mbox{i}}{a_f} \gamma_\mu \sin(a_fp_\mu) 
+ m^{-\epsilon_\alpha}(p)\right)P_{-\epsilon_\alpha}, 
\label{shift2}
\end{eqnarray}
where $f^{\epsilon_\alpha}_\mu$ and $m^{\epsilon_\alpha}(p)$,
$m^{-\epsilon_\alpha}(p)$ are left 
undetermined. If the action is invariant under both transformations, 
(\ref{S1}) and (\ref{S2}),
as was originally considered in ref.~15, then
\begin{equation}
m^{\epsilon_\alpha}(p) = m^{-\epsilon_\alpha}(p) = 
\frac{r}{a_f}\sum_\mu[1-\cos(a_fp_\mu)].
\label{shift3}
\end{equation}
Remember that the gauged (ungauged) fermion has chirality $\epsilon_\alpha$ 
($-\epsilon_\alpha$). The Dirac part of the action (\ref{LA}) is invariant 
under both transformations (\ref{S1}) and (\ref{S2}).
The Wilson term (\ref{PG}) is invariant under the transformation
(\ref{S1}), but not under (\ref{S2}). From the propagator (\ref{shift1}) we
read off that in the limit $a_f \rightarrow 0$ the theory 
describes a free, massless 
fermion with chirality $-\epsilon_\alpha$, plus an interacting fermion with 
chirality $\epsilon_\alpha$. The interacting fermion will, in general, 
require a mass counterterm to become massless, while for the free fermion 
no tuning of the mass is necessary. This means that the ungauged fermion
decouples. If the action is invariant under 
both transformations, then the interacting fermion will automatically be 
massless in the continuum limit, which follows from (\ref{shift3}). 
The gauged Wilson term of the vector model, e.g., is not invariant under any
of the two transformations.

We shall now compute the counterterm. In two dimensions the only diagrams 
one has to consider are those in fig.~1. In the continuum they require (at
most) a counterterm of the form
\begin{equation}
c\,e^2_\alpha \int \mbox{d}^2x A^2_\mu(x).
\end{equation}
To compute the coefficient $c$ we proceed as follows. The contribution of the
diagrams in fig.~1 to the effective action has the form
\begin{equation}
\sum_\alpha \frac{e^2_\alpha}{2} \int \mbox{d}^2k A_\mu(k)
\Pi_{\mu\nu}(k) A_\nu(-k),
\end{equation}
where $\Pi_{\mu\nu}(k)$ is the polarization tensor. The polarization
tensor is given by
\begin{equation}
\Pi_{\mu\nu}(k) = \left(\delta_{\mu\nu} A + \frac{(k_\mu+
\mbox{i}\epsilon_\alpha\widetilde{k}_\mu)
(k_\nu+\mbox{i}\epsilon_\alpha\widetilde{k}_\nu)}{k^2} B\right)
+ O(a_f),
\label{PT}
\end{equation} 
where $\widetilde{k}_\mu = \epsilon_{\mu\nu} k_\nu$. For the Wilson
term (\ref{PG}) the coefficients are
\begin{equation}
A = \int^\pi_{-\pi} \frac{\mbox{d}^2 p}{(2\pi)^2} \left\{ 
\frac{2c^2_1s^2_1 + 2rc_1s^2_1\hat{s}^2 + r^2s^2_1(\hat{s}^2)^2
-c^2_1 s^2}{(s^2+(\hat{s}^2)^2)^2} 
+ \frac{s^2_1 - rc_1\hat{s}^2}{s^2+(\hat{s}^2)^2}\right\}
\label{A}
\end{equation}
and 
\begin{eqnarray}
B &=& \int^\pi_{-\pi} \frac{\mbox{d}^2 p}{(2\pi)^2} 
\left\{ \frac{2c^2_1s^2_1 + 4rc_1s^2_1\hat{s}^2 
- r^2s^2_1(s^2 - (\hat{s}^2)^2) - c^2_1(s^2 
+(\hat{s}^2)^2)}{(s^2+(\hat{s}^2)^2)^2}\right. \nonumber \\
& & \hspace{2cm} + \left.\frac{s^2_1 - rc_1\hat{s}^2}
{s^2+(\hat{s}^2)^2}\right\},
\label{B}
\end{eqnarray}
where $c_\mu = \cos p_\mu$, $s_\mu = \sin p_\mu$ and 
$\hat{s}^2 = 2r \sum_\lambda \sin^2(p_\lambda/2)$. The integral 
in (\ref{B}) can be done analytically, giving
\begin{equation}
B = \frac{1}{4\pi}
\end{equation}
for all $r > 0$. For 
$r=0$ the result is $B = 1/\pi$, which is a factor of four larger, as it 
should be, because we have four identical flavors in this case. For (\ref{A})
and $r = 1$ we obtain numerically
\begin{equation}
A = 0.199006.
\end{equation}
The effective action, including the counterterm, is gauge invariant if
$A + 2c = B$. Thus we find for the coefficient of the counterterm
\begin{equation}
c = \frac{1}{2}(B-A) = -0.059714.
\label{coeff}
\end{equation}

\section{Numerical Test}

Because of lack of space we shall confine our tests to a few 
topics. Most important is the test of chiral gauge 
invariance of the effective action (\ref{effective}). If successful, it would 
mean that chiral gauge theories exist at the non-perturbative, constructive
level. Among the analytic results that need to be tested are factorization 
of the effective action and eq.~(\ref{imag}). These results are particularly 
important for numerical simulations of the theory. 

Before we begin our tests, let us briefly mention the extrapolation we are
using~\cite{GKSW}. We start from
the link variables on the original lattice,
\begin{equation}
U_\mu(s) \equiv \exp(\mbox{i}\theta_\mu(s)), \; -\pi < \theta_\mu(s) \leq \pi,
\end{equation}
where $s_\mu \in {\Bbb Z}$ are the lattice points on the original lattice. 
We define the plaquette
variables
\begin{equation}
P(s) = U_1(s) U_2(s+\hat{1}) U_1(s+\hat{2})^\dagger U_2(s)^\dagger \equiv 
\exp(\mbox{i}\omega(s)), \; -\pi < \omega(s) \leq \pi.
\end{equation}
We then have
\begin{equation}
\partial^+_1 \theta_2(s) - \partial^+_2 \theta_1(s) = \omega(s) - 2\pi n(s),
\end{equation}
with $n(s) = 0, \pm 1$. The cases $\theta_\mu(s) = \pi$, $\omega(s) = \pi$
correspond to exceptional configurations, for which the topological charge
is not defined. These configurations are of measure zero and will be 
excluded from our discussion. Following ref.~2 we then obtain for the 
continuum gauge field
\begin{eqnarray}
a A_1(x) &=& \theta_1(s) + [-\theta_1(s) + \theta_1(s+\hat{2}) 
- 2\pi n(s)](x_2-s_2) + 2\pi n(s) \theta(x_2-\bar{x}_2), \nonumber \\
a A_2(x) &=& \theta_2(s) + [-\theta_2(s) + \theta_2(s+\hat{1})](x_1-s_1) 
+ 2\pi n(s) (x_1-s_1)\delta(x_2-\bar{x}_2), \nonumber \\
\label{gf}
\\[-2.3em] \nonumber
\end{eqnarray}
where 
\begin{equation}
x \in c(s), \; c(s) = \{x\in {\Bbb T}^2 \mid s_\mu \leq x_\mu \leq s_\mu + 1\},
\end{equation}
and where $\bar{x}_2$ is implicitly defined by
\begin{equation}
\theta_1(s) + [-\theta_2(s) + \theta_2(s+\hat{1}) - \omega(s)] 
(\bar{x}_2 - s_2) = -\pi n(s).
\end{equation}
It follows that $-\pi \leq aA_\mu(x) \leq \pi$, except on the singular line
$x_2=\bar{x}_2$, $s_1 \leq x_1 \leq s_1+1$. The extrapolation is gauge
covariant, and the parallel transporters derived from the continuum gauge
field are consistent with those on the original lattice. In particular we have
\begin{equation}
a^2 F_{1 2}(x) = \omega(s), \; x \in c(s).
\end{equation}

We first discuss the case of trivial topology.
For compact lattice fields we have the decomposition
\begin{equation}
\theta_\mu(s) = \frac{2\pi}{L_\mu} t_\mu + \partial^+_\mu h(s) + 
\widetilde{\partial}^-_\mu f(s),
\end{equation}
where $f(s)$ is defined by $\square f(s) = \omega(s)$ with
$\square = \partial^+_\mu \partial^-_\mu$. Unlike in the continuum, where 
$t_\mu$ was determined up to an integer, $t_\mu$ is non-invariant under a 
wider class of large gauge transformations.

\begin{figure}[tbh]
%\vspace{ 2.0cm}
\begin{centering}
\epsfig{figure=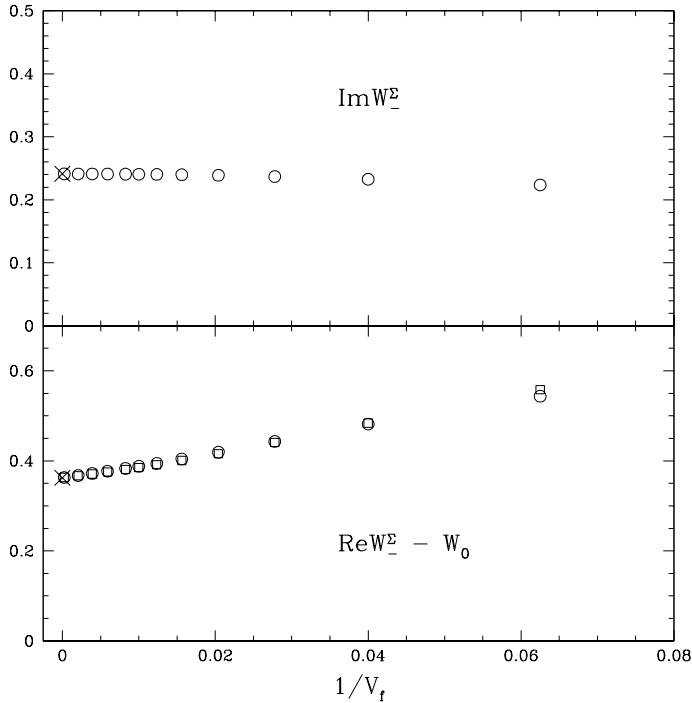,height=10.0cm,width=10.0cm}
%\vspace{-1.0cm}
\caption{The effective actions $\mbox{Re} W^\Sigma_- - W_0$ and 
$\mbox{Im} W_-$ ($\kreis$) for a pure toron field configuration 
with $(t,\phi) = (0.5,\pi/8)$ and ($e_\alpha,\epsilon_\alpha) 
= (1,-1)$ plotted against $1/V_f$. Also shown is $(W_V-W_0)/2$
($\square$). The numerical results are compared to the analytic
results ($\times$) given in (\ref{analy}), (\ref{imag}).}
\label{fig2}
\end{centering}
%\vspace{1.0cm} 
\end{figure}

All our calculations are for periodic boundary conditions for the 
gauge fields and anti-periodic boundary conditions for the fermions. We work
on symmetric lattices with $L_1 = L_2$. The lattice effective action is
computed by means of the Lanczos algorithm~\cite{Lanczos}. Note that 
the fermion matrix is not Hermitian in our case. The Lanczos vectors are 
re-orthogonalized after every iteration. Due to re-orthogonalization the 
CPU time grows with $V^3_f$, where $V_f = L^f_1 \times L^f_2$, and the 
memory demand grows with $V^2_f$.

To begin with, let us consider pure toron fields. Writing
\begin{equation}
t_1 = t \cos\phi, \; t_2 = t \sin\phi,
\end{equation}
we take
\begin{equation}
t = 0.5, \; \phi = \frac{\pi}{8}.
\label{tcfg}
\end{equation}
For this choice of $\phi$ the imaginary part of the effective action is 
close to its maximal value for any given value of $t$. As the anomaly is
zero in this case, we may consider a single chiral fermion. We choose 
$e_\alpha = 1$ and $\epsilon_\alpha = -1$. In fig.~2 we plot the  
effective action (\ref{WC}),
\begin{equation}
W^\Sigma_- = W_- + C = W_- + c \sum_{n, \mu} (2-U^f_\mu(n)
-U^f_\mu(n)^\dagger),
\end{equation}
with $c$ being given by (\ref{coeff}), as a function of $1/V_f$. We see 
that the 
numerical results converge to the analytic values in the limit $a_f
\rightarrow 0$. We furthermore see that $\mbox{Re}W^\Sigma_- \rightarrow
(W_V + W_0)/2$. 

\begin{figure}[t]
%\vspace{ 2.0cm}
\begin{centering}
\epsfig{figure=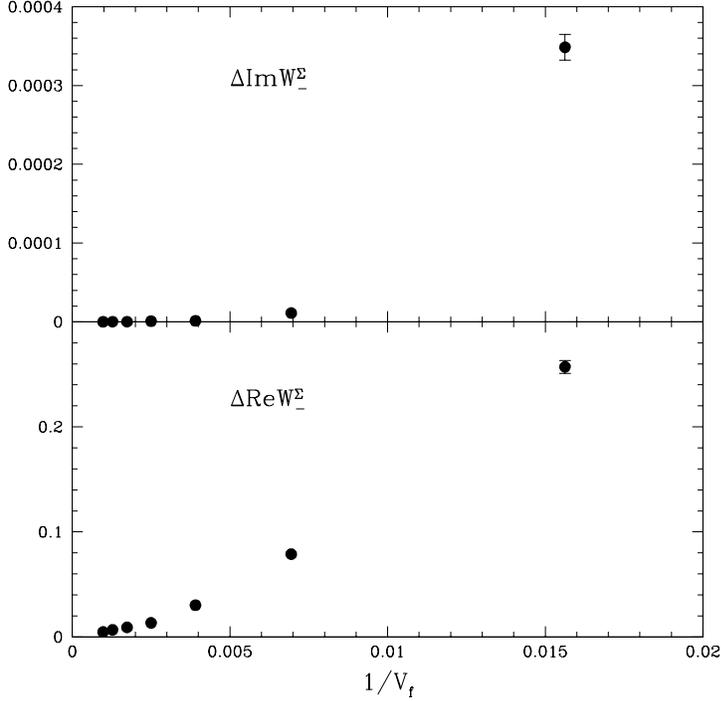,height=10.0cm,width=10.0cm}
%\vspace{-1.0cm}
\caption{The variance $\Delta \mbox{Re} W^\Sigma_-$ and 
$\Delta\mbox{Im} W_-$ for the toron field configuration 
plotted against $1/V_f$.}
\label{fig3}
\end{centering}
%\vspace{1.0cm} 
\end{figure}

Next we consider the effect of a small random lattice gauge transformation, 
described by $h(s)$, on the toron field configuration (\ref{tcfg}). To 
monitor the variation of the effective action 
under such transformations, we introduce the measure
\begin{equation}
\Delta X = \frac{1}{N} \sum_h |X^h - X|,
\end{equation}
where the sum is over a set of $N$ gauge transformations, $X$ is the starting 
value, and $X^h$ is the result after the gauge transformation. If the quantity
$X$ is gauge invariant, clearly $\Delta X = 0$. In fig.~3 we plot
$\Delta \mbox{Re} W^\Sigma_-$ and $\Delta \mbox{Im} W_-$ as a function
of $1/V_f$. The number of
gauge transformations $N$ varied between 20 and 100, depending on $V_f$.
We see that both $\Delta \mbox{Re} W^\Sigma_-$ and 
$\Delta \mbox{Im} W_-$ go to zero as expected in the limit 
$a_f \rightarrow 0$. 

\begin{figure}[t]
%\vspace{ 2.0cm}
\begin{centering}
\epsfig{figure=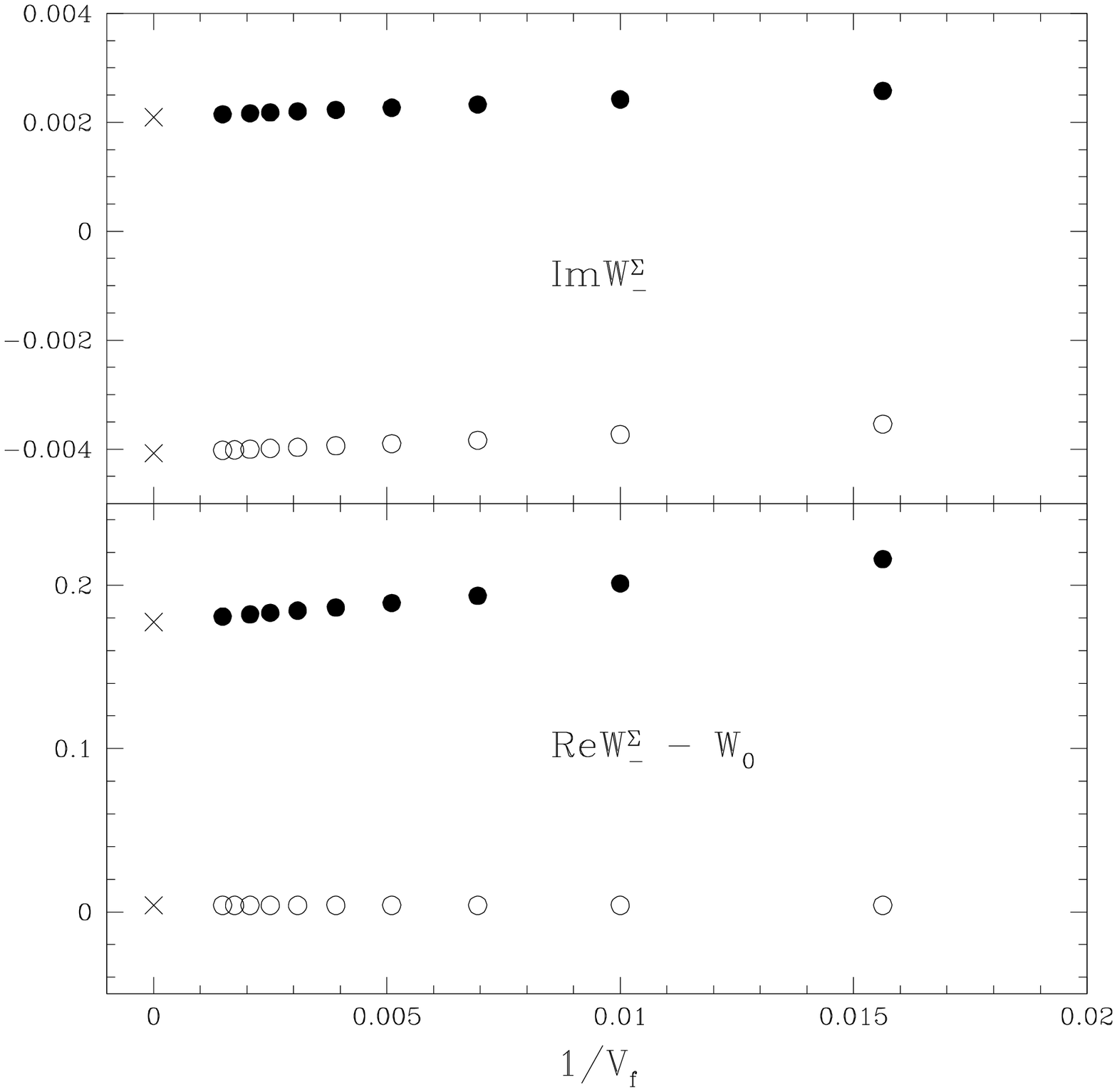,height=10.0cm,width=10.0cm}
%\vspace{-1.0cm}
\caption{The effective actions $\mbox{Re} W^\Sigma_- - W_0$ and 
$\mbox{Im} W_-$ for the configuration (\ref{mixed}) with 
$t_\mu = 0$ ($\kreis$) and $(t,\phi) = (0.5,\pi/400)$ ($\vollk$). 
The numerical results are compared to 
the analytic
results ($\times$) given in (\ref{analy}), (\ref{imag}).}
\label{fig4}
\end{centering}
%\vspace{1.0cm} 
\end{figure}

To check factorization (\ref{factor}) and the analytic results (\ref{analy}),
(\ref{imag}), 
one needs to consider more general gauge fields. We consider the continuum
gauge field
\begin{equation}
A_\mu(x) = c_\mu \cos \left(\frac{2\pi k_\mu x_\mu}{T_\mu}\right)
+ \frac{2\pi}{T_\mu} t_\mu,
\end{equation}
corresponding to a plane wave plus toron solution. On the lattice this gives
(cf. eq.~(\ref{link})) 
\begin{equation}
\theta^f_\mu(n) = a_f c_\mu \sin\left(\frac{\pi k_\mu}{L^f_\mu}\right)
\left(\frac{\pi k_\mu}{L^f_\mu}\right)^{-1} 
\cos\left(\frac{2\pi k_\mu n_\mu}{L^f_\mu} + \frac{\pi k_\mu}{L^f_\mu}\right)
+ \frac{2\pi}{L^f_\mu} t_\mu.
\label{mixed}
\end{equation}
The virtue of this configuration is that it allows to compute the continuum 
effective action analytically. We choose $T_1 c_1 = T_2 c_2 = 0.32$ and 
$k_1=1, k_2 = 0$. In fig.~4 we show $\mbox{Re}W^\Sigma_-$ and 
$\mbox{Im}W^\Sigma_-$ for $t_\mu = 0$ and $t=0.5, \phi=\pi/400$. The reason 
for choosing such a small value of $\phi$ was to make the plane wave and toron
contribution about equal in magnitude. We see in both cases that the numerical
results converge to the analytic values in the limit $a_f \rightarrow 0$.
In particular this confirms factorization, and 
that the non-toron part of the
effective action is given by the ${\Bbb R}^2$ result stated in eqs.~(\ref{W}) 
and (\ref{analy}).

We shall now consider a gauge field configuration generated by Monte-Carlo
with the pure gauge field action at $\beta = 1/e^2 = 6.0$. We have chosen a 
configuration with no vortex-antivortex pairs. In this case the imaginary
part of the effective action is not expected to be gauge invariant anymore
for a single chiral fermion. But it is expected to be gauge invariant in
the anomaly-free case. To test gauge invariance, we have computed
$\;\Delta\mbox{Re}W^\Sigma_-$, $\;\Delta\mbox{Im}W^\Sigma_-$ as well as
$\;\Delta\mbox{Im}W^\Sigma_a$, \hfill where the subscript $a$

\clearpage
\begin{figure}[tbh]
%\vspace{ 2.0cm}
\begin{centering}
\epsfig{figure=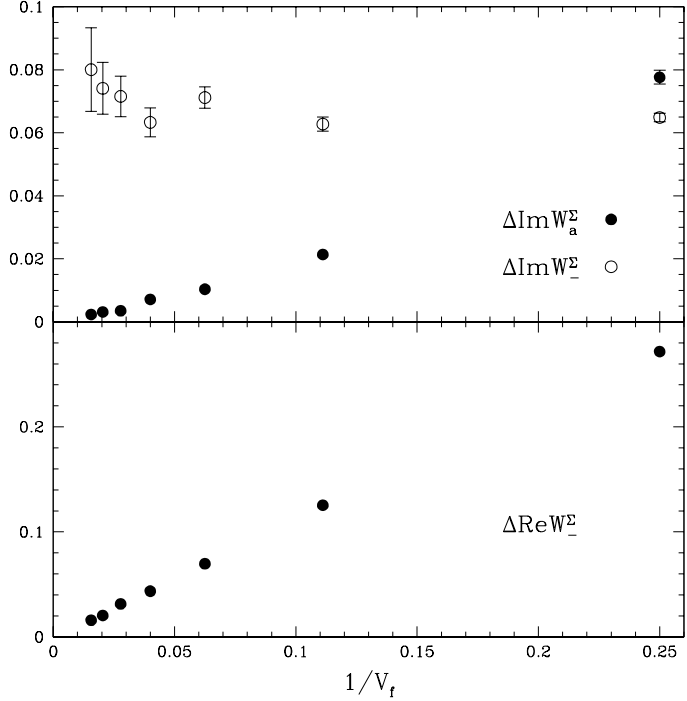,height=10.0cm,width=10.0cm}
%\vspace{-1.0cm}
\caption{The variance $\Delta \mbox{Re} W^\Sigma_-$ and 
$\Delta\mbox{Im} W_a$ ($\vollk$), together with 
$\Delta\mbox{Im} W_a$ ($\kreis$), for a Monte Carlo generated configuration
plotted against $1/V_f$.}
\label{fig5}
\end{centering}
%\vspace{1.0cm} 
\end{figure}

\noindent
 stands for the anomaly-free model with 
$\epsilon_\alpha e_\alpha = -1,-1,-1,-1,+2$. The result is shown in fig.~5.
We see that $\Delta\mbox{Re}W^\Sigma_-$ and $\Delta\mbox{Im}W^\Sigma_a$ go
to zero in the limit $a_f \rightarrow 0$, which means that gauge invariance
is restored. On the other hand, we find that $\Delta\mbox{Im}W^\Sigma_-$ does
not vanish, which is a consequence of the presence of the anomaly.

Let us now turn to topologically non-trivial configurations. Here we
restrict ourselves to the sector $|Q|=1$. For the lattice gauge field
$C^Q_\mu$ we take (cf. (\ref{gaugedef}))
\begin{equation}
C^{Q}_\mu(s) = 2\pi Q \epsilon_{\mu\nu} \partial^-_\nu G(s),
\label{eins}
\end{equation}
where $G(s)$ is the inverse lattice Laplacian. 
We have checked that the action (\ref{LA}) satisfies the index theorem. We 
have furthermore verified that the effective action factorizes into a toron
and a non-toron part, where both the real and imaginary part of the toron 
contribution are given by the analytic formulae (\ref{analy}) and (\ref{imag}).
We have no space to show the results here. We have also checked gauge 
invariance. In fig.~6 we show $\Delta \mbox{Re} W^\Sigma_-$ and 
$\Delta\mbox{Im} W_a$ for the configuration (\ref{eins}) and a single random 
gauge transformation. We have divided out $\mbox{Re} W^\Sigma_-$
because the action increases strongly with decreasing $1/V_f$. We find that
both $\mbox{Re} W^\Sigma_-$ and $\mbox{Im} W_a$ are gauge invariant in the
limit $a_f \rightarrow 0$.
\begin{figure}[tbh]
%\vspace{ 2.0cm}
\begin{centering}
\epsfig{figure=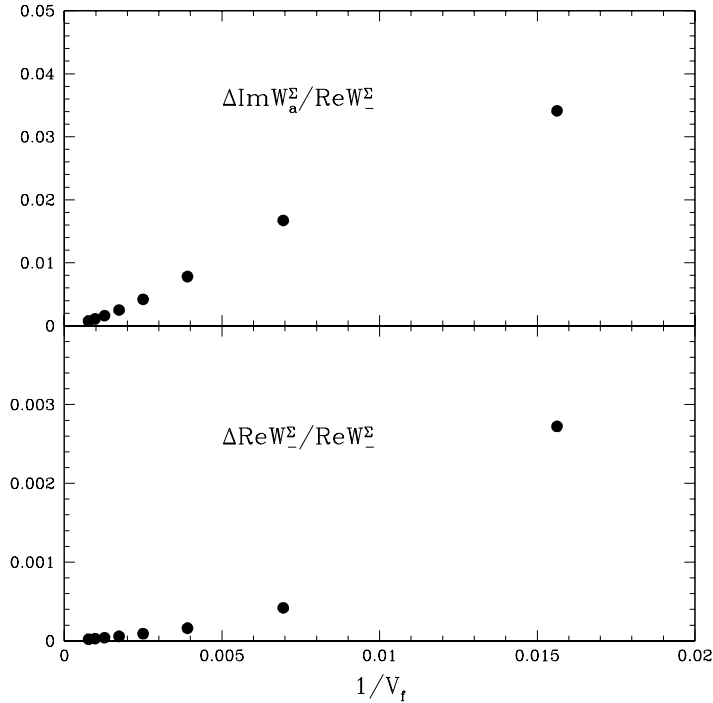,height=10.0cm,width=10.0cm}
%\vspace{-1.0cm}
\caption{The variance $\Delta \mbox{Re} W^\Sigma_-$ and 
$\Delta\mbox{Im} W_a$ for a $|Q|=1$ configuration plotted against $1/V_f$.}
\label{fig6}
\end{centering}
%\vspace{1.0cm} 
\end{figure}

\section{Conclusions}

Due to space limitations we were only able to show some of our results. 
In particular, we had to omit the discussion of large gauge
transformations. Some preliminary mention of this topic can be found 
in ref.~19. In general, the action is not invariant under such
gauge transformations. The action of the transformed field will, however,
diverge in the  
limit $a_f \rightarrow 0$, so that this configuration has zero 
weight in the partition function. This is not a shortcoming of our method. 
The same result is found in the continuum. 

We may conclude that we have found a non-perturbative, gauge invariant
formulation of chiral gauge theories. In the sector of trivial topology
the real part of the effective action converges to $(W_V + W_0)/2$ in the
continuum limit, whereas the imaginary part becomes a function of the toron 
field only in the anomaly-free model. The effective action
may then be written
\begin{equation}
\widehat{W} = \frac{1}{2}(W_V + W_0) 
+ \mbox{i}\, \mbox{Im}W,  
\end{equation}
with
\begin{equation}
\mbox{Im} W \equiv \mbox{Im} W(a), \; a_\mu = \frac{2\pi}{T_\mu} t_\mu
\end{equation}
being given by (\ref{imag}).
The result for the real part, namely that it is given by half the action of 
the vector theory, was already conjectured in ref.~13. 
The imaginary part of the effective action can be computed analytically
directly from the lattice gauge field. Thus we have arrived at an action which
can be simulated on the original lattice with not much more effort than that
of the vector theory.  

In the sector of non-trivial topology the result for the real part has to 
be modified, while the result for the imaginary part will be unchanged.
We shall return to this problem in the near future. 

We see no problem in extending the method to higher dimensions. Work on a 
chiral U(1) model in four dimensions is in progress.

\section*{Acknowledgement}

One of us (G.S.) thanks M. G\"ockeler for discussions on the subject of 
this talk. He also likes to thank T. Suzuki for his kind hospitality at
Kyoto as well as JSPS for financial support.

\end{document}